\documentclass[sigconf]{acmart}
\AtBeginDocument{%
  \providecommand\BibTeX{{%
    \normalfont B\kern-0.5em{\scshape i\kern-0.25em b}\kern-0.8em\TeX}}}

\copyrightyear{2021}
\acmYear{2021}
\setcopyright{acmlicensed}

\acmConference[CIKM '21]{Proceedings of the 30th ACM International Conference on Information and Knowledge Management}{November 1--5, 2021}{Virtual Event, QLD, Australia.}
\acmBooktitle{Proceedings of the 30th ACM International Conference on Information and Knowledge Management (CIKM '21), November 1--5, 2021, Virtual Event, QLD, Australia}
\acmPrice{15.00}
\acmDOI{10.1145/3459637.3481903}
\acmISBN{978-1-4503-8446-9/21/11}
\usepackage{xcolor}

\usepackage{amsmath,graphicx}
\usepackage{threeparttable}
\usepackage{multirow}
\usepackage{subfigure}
\usepackage{amsmath,amsfonts}
\usepackage{algorithmic}
\usepackage{graphicx}
\usepackage{textcomp}

\usepackage[linesnumbered, ruled, vlined]{algorithm2e}

\begin{document}
\fancyhead{}

\title{CloudRCA: A Root Cause Analysis Framework \\ for Cloud Computing Platforms}



\author{
   Yingying Zhang, Zhengxiong Guan, Huajie Qian, Leili Xu, Hengbo Liu, Qingsong Wen,\\ Liang Sun, Junwei Jiang, Lunting Fan, Min Ke
}
\affiliation{
    \institution{Alibaba Group}
    \city{Hangzhou}
    \country{China}
}
\email{congrong.zyy, zhengxiong.gzx, h.qian, leili.xll, yangyang.lhb, qingsong.wen, liang.sun, junwei.jjw@alibaba-inc.com, lunting.fan, dawu@taobao.com}
\renewcommand{\shortauthors}{Zhang et al.}


\begin{abstract}
As business of Alibaba expands across the world among various industries, higher standards are imposed on the service quality and reliability of big data cloud computing platforms which constitute the infrastructure of Alibaba Cloud. However, root cause analysis in these platforms is non-trivial due to the complicated system architecture. In this paper, we propose a root cause analysis framework called CloudRCA which makes use of heterogeneous multi-source data including Key Performance Indicators (KPIs), logs, as well as topology, and extracts important features via state-of-the-art anomaly detection and log analysis techniques. The engineered features are then utilized in a Knowledge-informed Hierarchical Bayesian Network (KHBN) model to infer root causes with high accuracy and efficiency. Ablation study and comprehensive experimental comparisons demonstrate that, compared to existing frameworks, CloudRCA 1) consistently outperforms existing approaches in f1-score across different cloud systems; 2) can handle novel types of root causes thanks to the hierarchical structure of KHBN; 3) performs more robustly with respect to algorithmic configurations; and 4) scales more favorably in the data and feature sizes. Experiments also show that a cross-platform transfer learning mechanism can be adopted to further improve the accuracy by more than 10\%. CloudRCA has been integrated into the diagnosis system of Alibaba Cloud and employed in three typical cloud computing platforms including MaxCompute, Realtime Compute and Hologres. It saves Site Reliability Engineers (SREs) more than $20\%$ in the time spent on resolving failures in the past twelve months and improves service reliability significantly.
\end{abstract}
%








\maketitle

\section{Introduction}
Nowadays, tens of millions of products and billions of consumers are involved in Alibaba's e-commerce platforms. A tremendous amount of data at PB-level is processed every day on Alibaba's big data cloud computing platforms, in order to complete transactions and provide appropriate recommendations to customers. The number of merchants, consumers, daily processed data still keep increasing and reach its peak on the Double 11 global shopping festival every year. Therefore, stability and fast fault recovery of cloud platforms are essential for supporting these business scenarios. 

The main challenges of root cause identification for big data cloud computing platforms at Alibaba include: (1) The ever increasing scale and complexity of computing platforms as well as multi-source, large-volume and unstructured data make it difficult for SREs to perform troubleshooting manually; (2) Real-time business scenarios are ubiquitous and time-sensitive, hence the whole process of detection and diagnosis needs to be finished in a faster way, for which efficiency is usually measured by MTTR (Mean Time to Resolve); (3) Faults are relatively rare in those systems, which limit the amount of samples available for training models; (4) Due to the diversity of Alibaba's business scenarios, there are different kinds of computing platforms serving various requests, such as large-scale batch computing, real-time computing, interactive analysis, etc. Therefore the root cause analysis workflow needs to be widely applicable to computing platforms with different architectures.

There are several well-studied root cause analysis (RCA) methods that have been demonstrated to perform competitively, such as CloudRanger proposed in \cite{wang2018cloudranger} from IBM, OM knowledge graph based method (O\&M Graph for short) proposed by \cite{qiu2020causality}, and LogCluster proposed by \cite{lin2016log} from Microsoft. These methods utilize either KPI time series or logs for root cause analysis, however, none of them uses these two important sources of data at the same time. Moreover, none of these methods provides direct root cause types which can guide SREs or automatic tools to take actions and recover the systems without further manual investigation. Specifically, in CloudRanger and O\&M Graph, the root cause inference result is the KPI node on the graph, while for LogCluster, the inference result is the representative log sequences.

In this paper, we propose a root cause analysis workflow called CloudRCA, and the main contributions of our study are: (1) The framework integrates multiple sources of data including not only monitoring metrics but also system logs together with expert knowledge, resulting in higher accuracy compared with other methods that only consider KPIs or logs. We also improve the algorithms used in CloudRCA, such as optimizing the pruning strategy in log template extraction, improving log clustering through Natural Language Processing feature representations and defining the hierarchical root cause layers based on traditional Bayesian networks, all leading to higher accuracy and efficiency of CloudRCA than existing RCA methods. This has been demonstrated through comprehensive experiments in our study. (2) CloudRCA is able to deal with brand new types of root causes that have never been seen before, thanks to the innovative design of hierarchical root cause layer in the Bayesian network. This improvement in Bayesian network enables the model to infer the most likely root cause module even when encountered with new types. (3) CloudRCA is a general framework that can be easily deployed on various cloud systems with minor adaptations. It has been implemented on Alibaba's three big data cloud computing platforms in practice, namely MaxCompute, Realtime Compute and Hologres, and helps SREs reduce the time of resolving failures by more than 20\% in the past twelve months. (4) A comprehensive set of experiments are conducted, from which key insights regarding RCA modeling are derived.

\section{Related work}

Many research papers on RCA focus on complex large-scale systems~\cite{qiu2020causality,thalheim2017sieve,weng2018root}. They can be grouped into the following categories:

\subsubsection{Log-based methods}

The common methodology to RCA is to analyze log files to identify problems that occurred in the system. The problems are then examined to identify potential causes. \cite{lin2016automated} introduced a classic log mining method to diagnose and locate anomalies in distributed systems. According to \cite{xu2017logdc,jia2017logsed}, the log mining method also plays an important role in the RCA of cloud applications. However, the limited coverage of log features makes it difficult for methods relying only on log features to achieve high accuracy.


\subsubsection{Correlation analysis-based methods}  

These methods correlate events, metrics and alerts to locate the root cause of an anomaly or failure. \cite{zeng2014mining} proposed a parametric model to describe noisy time lags from fluctuating events. \cite{marvasti2013anomaly} introduced a model of statistical inference to manage complex infrastructures based on their anomalous events data obtained from an intelligent monitoring engine. But without considering the topology of infrastructures and other domain knowledge, the correlated results may be hard for SREs to understand.

\subsubsection{DAG-based methods}  

Deploying applications in cloud environments has become a tendency. As calls between systems become more and more complex, it is difficult for SREs to directly locate root causes. In the RCA research for cloud applications, the most representative are related studies from CloudRanger~\cite{wang2018cloudranger} from IBM and OM knowledge graph based method~\cite{qiu2020causality}. These methods build a dependency or causality graph from historical KPIs data and use the constructed graph to infer root causes online. But those works don't make full use of expert knowledge and multiple sources of data including logs.

\section{Proposed CloudRCA Framework}
\begin{figure}[h]
\centerline{\includegraphics[scale=0.35]{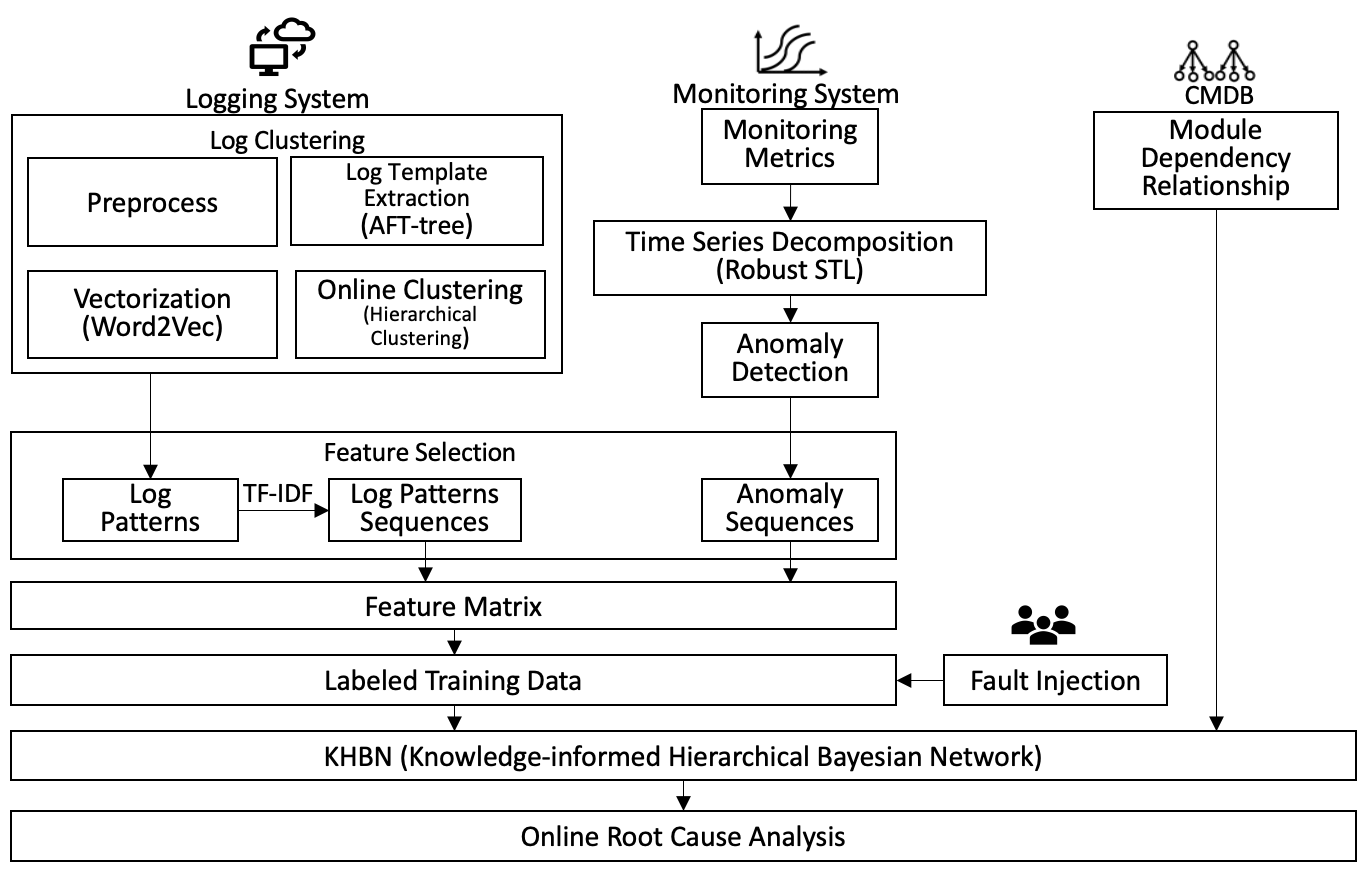}}
\caption{The workflow of the proposed CloudRCA framework.}
\label{framework}
\end{figure}

Our framework features a novel integration of three sources of information, including metrics from monitoring system, log messages from logging system, and module dependency relationship in Configuration Management Database (CMDB). An overview is shown in Figure \ref{framework}. First, in the feature engineering stage, data from multiple sources are converted into unified feature matrices through time series anomaly detection and log clustering algorithms. Then a KHBN(Knowledge-informed Hierarchical Bayesian Network), combining processed signal data and domain knowledge of module dependency, is built to infer root causes in real time. Next we elaborate on each component of the framework.

\subsection{Metrics Anomaly Detection}\label{sec: anomaly detection}
This subsection presents the extraction of anomalous metrics collected from monitoring systems as a basis for RCA. Based on the analysis of historical failure data in our cloud computing platforms, we summarize four kinds of typical anomalies that are frequently related to system breakdowns, including spikes and dips, change of mean, change of variance, and long-term trend. We design a decomposition-based scheme to detect these anomalies.

We first identify period lengths of the time-series metrics. Since metrics in stream processing typically exhibit daily and/or weekly periodicity, we design a simplified RobustPeriod algorithm~\cite{WenRobustPeriod20} which adopts wavelet transform to isolate single periodicity and then identifies the exact period from the peaks of the auto-correlation function (ACF). Next, we decompose the time series into tread, seasonality, and remainder components. Traditional STL (Seasonal-Trend decomposition using Loess) proposed in~\cite{robert1990stl} suffers from less flexibility when seasonality period is long and is vulnerable to noises. Therefore we use a novel and robust decomposition method called RobustSTL~\cite{wen2019robuststl,FastRobustSTL_wen2020,yang2021robust}. RobustSTL first extracts the trend component by solving a regression problem using the least absolute deviation loss with sparse regularization, then estimates the seasonality component with a non-local seasonal filtering.



With the decomposed time series, we identify those anomalies  in a ``divide-and-conquer'' manner, as summarized in Table~\ref{tab:component-stat-tests}. The main idea is to detect different types of anomalies from different components using proper statistical tests. Furthermore, two improvements are adopted for practical considerations. First, robust statistics~\cite{zoubir2012robust}, such as median and median absolute deviation~\cite{leys2013detecting} in place of simple mean and variance~\cite{hochenbaum2017automatic}, are utilized in the statistical tests to hedge against outliers and noises in time series. Second, online versions of these tests are implemented by updating statistics incrementally via a bisection algorithm~\cite{ali2017n}, which greatly boosts the computational speed.




\begin{table}[h]
\small
\begin{center}
  \caption{\small Summary of the designed decomposition-based anomaly detection scheme.}
  \begin{tabular}{ c | c | c }
    \hline
    Anomaly Type & Component & Statistical Test \\ \hline 
    Spikes \& Dips & remainder & Extreme studentized deviate test \\\hline
    Change of Variance & remainder & F-test \\\hline
    Change of Mean & trend & T-test \\\hline
    Long-term Trend & trend &  Mann-Kendall test \\\hline 
  \end{tabular} 
\label{tab:component-stat-tests}
\end{center}
\end{table}

\subsection{Log Templates Extraction and Clustering}\label{sec: log clustering}
Apart from monitoring metrics, numerous logs are consistently produced by many different modules that constitute a large-scale cloud computing platform. The sheer volume, varying syntax and semantics across modules make log analysis a non-trivial task. This section describes an effective template extraction method and pattern clustering that aggregates the vast raw log messages into several representative patterns.

After a standard log preprocessing process, including stemming and case-folding, we extract templates from message contents by removing variables such as IP address, table name, interface IDs, etc., that are not critical in diagnosis.

We develop an incrementally trainable algorithm called Adaptive Frequent Template tree (AFT-tree) based on FT-tree \cite{zhang2018prefix} to obtain log templates automatically instead of using regular expression. AFT-tree is designed with better efficiency and adaptivity. First, we use dictionaries to store children of nodes instead of lists, and avoid duplicated searching while inserting nodes. Our approach turns out $27$ times faster than FT-tree on our datasets. Second, unlike FT-tree that prunes the tree based on the number of children, AFT-tree uses the number of leaves of a node for pruning so that the template length can adapt to the length of the log to capture important non-variables.


After extracting the structural templates of the logs, we cluster them by the semantics with NLP algorithms. We apply Word2vec \cite{mikolov2013efficient,mikolov2013distributed} to obtained vector-representations of the logs. Then we perform hierarchical clustering \cite{gower1969minimum} based on the cosine similarity between the vectors to aggregate logs into different clusters, which we call log patterns.


Since logs are constantly generated, our clustering algorithm needs to run in an efficient online mode. To this end, we first conduct an offline training to obtain the clusters. For each cluster, we select a representative log by choosing the center of the cluster. To be precise, we compute a score for each log in a cluster based on its average distance to other logs within the same cluster:
\begin{equation*}
\textrm{Score}(i)=\frac{1}{n-1} \sum_{j=1}^{n}(1-\textrm{similarity}(\text{log}_i,\text{log}_j))
\end{equation*}
where $n$ is the total number of logs in the cluster, and the log content with the minimum score is selected as the representative. Then in the online mode, new logs are clustered together with the extracted representatives, where they either get distributed into the existing clusters or form new ones by their own. To be more specific, when a new log $s$ comes, we compute the similarity of the input log $s$ with the representatives of existing clusters. If the similarity is above a threshold $\theta$, we assign the input log message to the cluster with the greatest similarity; otherwise we create a new cluster for $s$, with $s$ itself as the representative. In practice, the threshold can be calculated as:
\begin{equation*}
\textrm{$\theta$}=\frac{1}{m} \sum_{i=1}^{m}\textrm{average\_similarity\_cluster}_{i}
\end{equation*}
where $m$ is total number of clusters, and $\textrm{average\_similarity\_cluster}_{i}$ is the average similarity of the representative log of cluster $i$ with other logs in cluster $i$. 

\subsection{Knowledge-informed Hierarchical Bayesian Network (KHBN)}\label{sec: expert knowledge}

In addition to logs and monitoring metrics, we also have tree-based data stored in Configuration Management Database (CMDB) which describe complicated dependency relationships between modules and root cause types that are useful for issue-tracking but hard to infer otherwise. All these are integrated into RCA through a hierarchical Bayesian network.




To prepare for network construction, we transform the logging and monitoring data into standard feature data. We build time series for log patterns indicating whether a certain log pattern appears (marked $1$) or not (marked $0$). Similarly, each monitoring item is marked $1$ if it's deemed as an anomaly by the algorithm and $0$ otherwise. For convenience, both transformed series are called metrics in our network. Some metrics may not be valuable for root cause identification and adversely affect the model, therefore we calculate the TF-IDF score of each feature to select only the most informative ones.


\begin{figure}[h]
\centering
\includegraphics[scale=0.4]{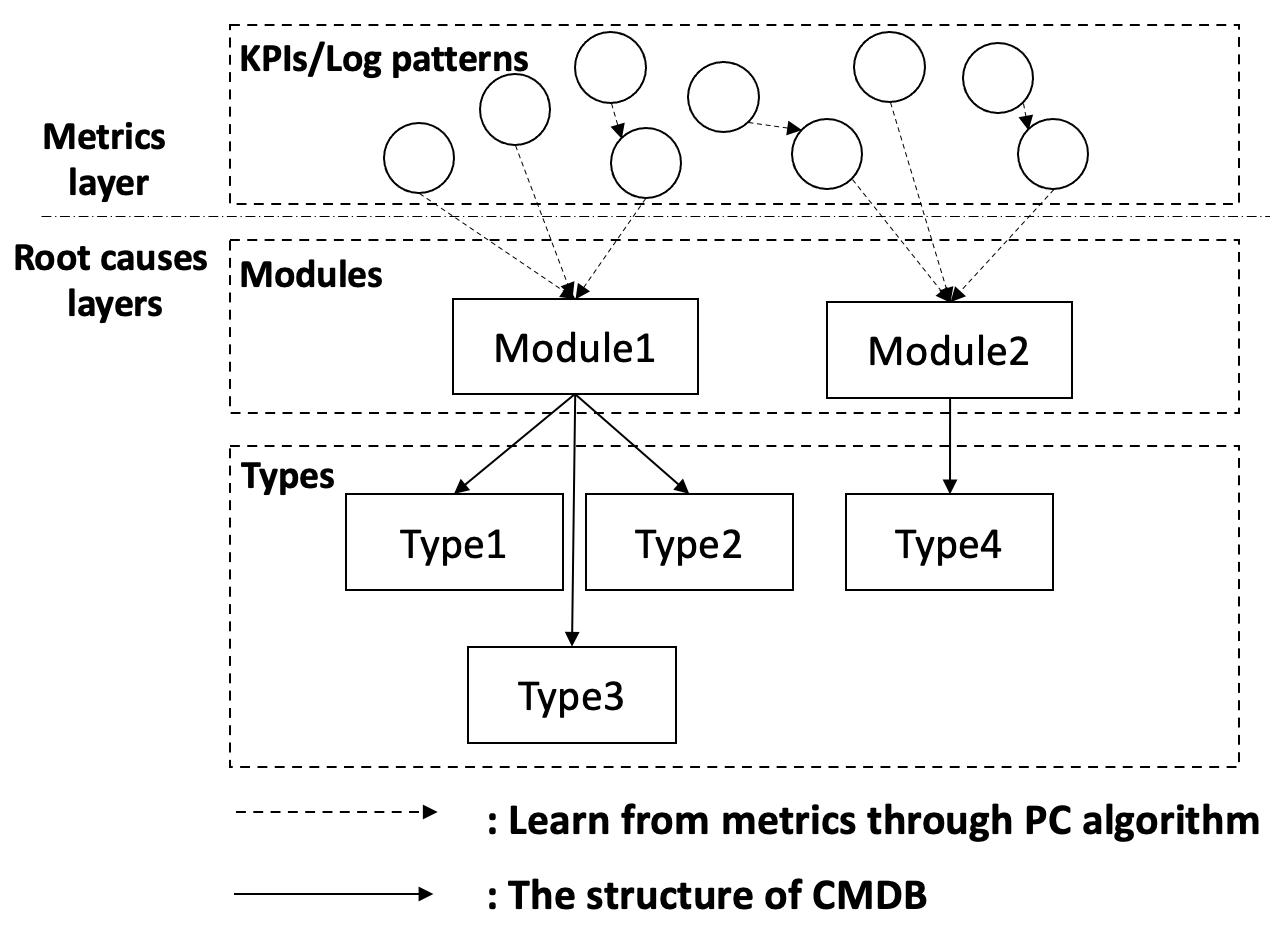}
\caption{The structure of the proposed KHBN diagram.}
\label{network}
\end{figure}
The structure of KHBN is shown in Figure \ref{network}. The nodes of the first layer are metrics selected from logging and monitoring, and root cause layer is composed of module-level nodes as well as type-level nodes. We build the network structure in two phases, an allocation phase and a causal learning phase. In the allocation phase, we use topology information from CMDB to generate the initial edges between nodes. In the causal learning phrase, directed edges that represent causal relationship between nodes are constructed based on system metrics collected from both normal and abnormal periods. This is achieved by the PC algorithm \cite{wang2018cloudranger}, a standard method for estimating causal structures. After constructing the network structure, we calculate the conditional probability table of each node using the maximum likelihood estimate. In this way we obtain a complete Bayesian network that incorporates CMDB topology and system historical metrics. The network can be updated regularly as new data are collected.
We then identify the root cause type through exact inference. When system fails, we input those feature data into KHBN model we trained with historical data, and select the node from the root cause types layer with the highest probability as the root cause, which is defined as
\begin{equation*}
t_{i}^{*}=\textrm{argmax}_{t_{i}}\; P(t_{i}|m_{k},s_{1},s_{2},...,s_{j})P(m_{k}|s_{1},s_{2},...,s_{j})    
\end{equation*}
where $t_{i}$ stands for a certain type of root cause, $m_{k}$ is the node in the module layer that points to $t_{i}$, and $s_{1},\ldots, s_{j}$ represent the observed nodes in the first layer. $t_{i}^{*}$ is the root cause type inferred by our model which has the highest conditional probability.



\section{Real-World Deployments}

Our root cause localization workflow is deployed in several big data cloud computing platforms of Alibaba, including MaxCompute, Realtime Compute and Hologres. The evaluation starts with the illustration of the importance of each step in the workflow, followed by the comparison of the algorithms we take in each phase with other state-of-art algorithms, and we also compare and analyze the performance on those three platforms. We then demonstrate the advantage of CloudRCA in dealing with brand new types of root causes, analyze the effect of eliminating predefined knowledge and assess the effectiveness of transfer learning in our workflow. Sensitivity study and overhead analysis are also performed. In Section \ref{sec:case study}, we share three real-world cases during the employment of CloudRCA on three platforms. 

\subsection{Cloud Computing Platforms at Alibaba}

\subsubsection{MaxCompute}  

MaxCompute is a data processing platform for large-scale data warehousing. It has been widely used and trusted by Alibaba and other customers, and is responsible for various large-scale data analysis scenarios in e-commerce, security, manufacturing and logistics. Currently, MaxCompute has more than 100 thousand servers located in data centers around the world, and can process more than one million jobs each day. The stored data has reached EB-level data volume long before, making it one of the leading products in the global market. 
The system architecture of MaxCompute is very similar to those of other big data computing engines in the industry. The most underlying is the infrastructure including host, network and so on. The storage module is a hierarchical storage distributed file system called Pangu. The resource management module is a scheduler with decentralized multi-scheduler architecture, which was designed to schedule large-scale distributed resources and improve cluster utilization.

\subsubsection{Realtime Compute}  

Realtime Compute offers a one-stop, high-performance platform that enables real-time big data processing based on Apache Flink. It not only provides internal services at Alibaba, but also supports Flink-based cloud products in the entire developer ecosystem by using cloud product APIs at Alibaba Cloud, and has become a popular stream computing engine even the de facto standard in the real-time computing industry domestically and internationally. The number of Alibaba Cloud's real-time computing jobs has reached more than 35,000 with the computing scale over 1.5 million CPUs, and the peak of real-time data processing during this year's Double 11 reached 4 billion records per second.
Flink computing engine runs on open-source Hadoop clusters. YARN of Hadoop is used for resource management and scheduling, and HDFS is used for data storage.

\subsubsection{Hologres}

Hologres is a cloud-native Hybrid Serving \& Analytical Processing (HSAP) system that is seamlessly integrated with the big data ecosystem. It integrates the processes of analytical processing and knowledge serving and supports highly concurrent writes and queries at a speed of up to 100 million transactions per second (TPS). Hologres takes a cloud-native design architecture where the computation and storage layers are decoupled. The computing module is deployed on Kubernetes while the storage layer is Pangu by default, which is a high performance distributed file system, similar to MaxCompute.


The three platforms above are typical representatives of the big data ecosystem at Alibaba. They serve different purposes of data processing and analysis and vary in scale. Despite the diversity in compute engine, scheduler and storage module design, they share some infrastructure such as host, network, and their basic system modules are similar. Table \ref{System Modules and Corresponding Root Cause Types of the Three Platforms} shows the modules and corresponding root cause types of the three platforms. It's worth mentioning that all the three platforms have a trend toward the cloud-native data computing service, which means that they may share more infrastructure in storage and resource scheduler in the near future. The main differences among the three platforms are summarized in Table \ref{Main Differences among the Three Platforms}.

\begin{table*}[h]
\footnotesize
\caption{System Modules and Corresponding Root Cause Types of the Three Platforms}
\begin{tabular}{|c|c|c|c|}
\hline
\multirow{2}{*}{Module} & Realtime Compute & MaxCompute & Hologres \\ \cline{2-4} 
 & \multicolumn{3}{c|}{Root Cause Types} \\ \hline
\multirow{4}{*}{Resource Scheduler} & YARN NM decommissioned & Fuxi master fail & ASI server overload \\ \cline{2-4} 
 & YARN RM switch & Fuxi tobo fail & ASI node fail \\ \cline{2-4} 
 & YARN resource preemption & Fuxi apiserver overload & ASI apiserver overload \\ \cline{2-4} 
 & … & … & … \\ \hline
\multirow{6}{*}{Storage} & HDFS service unavailable & \multicolumn{2}{c|}{pangu server unavailable} \\ \cline{2-4} 
 & HDFS usage over limit & \multicolumn{2}{c|}{pangu master failover} \\ \cline{2-4} 
 & HDFS call queue full & \multicolumn{2}{c|}{pangu master queue size full} \\ \cline{2-4} 
 & … & \multicolumn{2}{c|}{pangu server write slow} \\ \cline{2-4} 
 &  & \multicolumn{2}{c|}{pangu chunkserver failover} \\ \cline{2-4} 
 &  & \multicolumn{2}{c|}{…} \\ \hline
\multirow{6}{*}{Host} & \multicolumn{3}{c|}{oom} \\ \cline{2-4} 
 & \multicolumn{3}{c|}{io hang} \\ \cline{2-4} 
 & \multicolumn{3}{c|}{disk failure} \\ \cline{2-4} 
 & \multicolumn{3}{c|}{cpu usage over limit} \\ \cline{2-4} 
 & \multicolumn{3}{c|}{machine breakdown} \\ \cline{2-4} 
 & \multicolumn{3}{c|}{…} \\ \hline
\multirow{4}{*}{Network} & \multicolumn{3}{c|}{Martnet exception} \\ \cline{2-4} 
 & \multicolumn{3}{c|}{QoS exception} \\ \cline{2-4} 
 & \multicolumn{3}{c|}{LVS exception} \\ \cline{2-4} 
 & \multicolumn{3}{c|}{…} \\ \hline
\multirow{3}{*}{Other} & Upstream-TT & Tunnel & POP \\ \cline{2-4} 
 & Upstream-SLS & Frontend & DNS \\ \cline{2-4} 
 & … & … & … \\ \hline
\end{tabular}
\label{System Modules and Corresponding Root Cause Types of the Three Platforms}
\end{table*}

\begin{table*}[h]
\footnotesize
\caption{Comparison of the Three Platforms}
\begin{tabular}{|c|c|c|c|}
\hline
 Module & RealtimeCompute & MaxCompute & Hologres \\ \hline
feature of compute engine & batch & stream & Hybrid Serving \& Analytical Processing \\ \hline
cluster scale & \multicolumn{3}{c|}{MaxCompute \textgreater RealtimeCompute \textgreater Hologres} \\ \hline
resource scheduler & YARN & Fuxi & ASI \\ \hline
storage & HDFS & Pangu & Pangu \\ \hline
\end{tabular}
\label{Main Differences among the Three Platforms}
\end{table*}

\subsection{Dataset and Setup}

Related data over the past five years, including time series metrics, logs and tree-based data from CMDB, are collected for each platform. A summary of the data is shown in Table \ref{The Datasets of the three cloud computing platforms}.  We extract positive samples when the system is running normally and negative samples when the systems fail, including real faults and injected faults. All the positive samples are used as training set, which can help learn some correlation among those metrics, while negative samples are split into training set and test set with 60-40 ratio. As we can see from the Table, MaxCompute has the most negative samples due to its large scale and long-term data accumulation, while the dataset of Hologres are relatively small since it's a fairly new product.

To evaluate the performance of CloudRCA and other root cause analysis models, we use precision, cover rate, and f1-score, which are defined as:
\begin{equation*}
precision=\sum_{i}^{n} precision\_of\_each\_type(i)/n
\end{equation*}
\begin{equation*}
cover\_rate=covered\_types/all\_types
\end{equation*}
\begin{equation*}
f1\_score=2*precision*cover\_rate/(precision+cover\_rate)
\end{equation*}
precision\_of\_each\_type(i) represents the percentage of samples that are correctly predicted for a certain root cause type i. The precision is the average of all types' precision. It's worth mentioning that a type is called covered when $\geq 60\%$ of testing samples in this type are correctly predicted. cover\_rate indicates the covered types out of all types. f1\_score is a composite indicator that balances precision and cover\_rate. 

Our workflow is parallel, and all experiments are run on up to four machines, each with 16 Intel Xeon E5-2650 @2.00GHz cores and 128GB RAM, running Debian GNU/Linux9.4.

\begin{table}[h]
\footnotesize
\caption{The Dataset of the three cloud computing platforms}
\begin{tabular}{|c|c|c|c|c|}
\hline
\multirow{2}{*}{Production} & \multirow{2}{*}{Module} & \multicolumn{2}{c|}{\begin{tabular}[c]{@{}c@{}} Training \\ Set \end{tabular}} & \begin{tabular}[c]{@{}c@{}} Testing \\ Set \end{tabular} \\ \cline{3-5} 
 &  & \begin{tabular}[c]{@{}c@{}} Positive \\ Samples \end{tabular} & \begin{tabular}[c]{@{}c@{}} Negative \\ Samples \end{tabular} & \begin{tabular}[c]{@{}c@{}} Negative \\ Samples \end{tabular} \\ \hline
\multirow{5}{*}{MaxCompute} & Storage & 861 & 123 & 82 \\ \cline{2-5} 
 & Resource Scheduler & 924 & 132 & 88 \\ \cline{2-5} 
 & Host & 798 & 114 & 76 \\ \cline{2-5} 
 & Network & 777 & 111 & 74 \\ \cline{2-5} 
 & Other & 840 & 120 & 80 \\ \hline
\multirow{5}{*}{Realtime Compute} & Storage & 273 & 39 & 26 \\ \cline{2-5} 
 & Resource Scheduler & 287 & 41 & 27 \\ \cline{2-5} 
 & Host & 455 & 65 & 43 \\ \cline{2-5} 
 & Network & 378 & 54 & 36 \\ \cline{2-5} 
 & Other & 189 & 27 & 18 \\ \hline
\multirow{5}{*}{Hologres} & Storage & 140 & 20 & 13 \\ \cline{2-5} 
 & Resource Scheduler & 168 & 24 & 16 \\ \cline{2-5} 
 & Host & 252 & 36 & 24 \\ \cline{2-5} 
 & Network & 182 & 26 & 17 \\ \cline{2-5} 
 & Other & 105 & 15 & 10 \\ \hline
\end{tabular}
\label{The Datasets of the three cloud computing platforms}
\end{table}

\subsection{Ablation Study}
We design experiments to study the contribution of different components of the proposed framework to the overall performance. We also compare the techniques we used at each stage with their alternatives to demonstrate their strengths.

\subsubsection{Validation of Feature Engineering}


The feature engineering stage in our workflow combines multi-source data and extracts the most relevant features. To demonstrate its effectiveness, we conduct experiments on each of the three platforms. For each step (i.e., anomaly detection, log template extraction, and log clustering) in feature engineering, we investigate two settings where: 1) the exact proposed method is used;  2) the step is not performed at all. Results are shown in Table \ref{Ablation analysis}. Note that ``NONE'' in the table stands for the setting where the particular step is skipped. For example, in the 3rd row of Table \ref{Ablation analysis} anomaly detection is not performed, and the original kpi metrics are directly used to build KHBN, the root cause inference network.


\begin{table*}[h]
\footnotesize
\caption{Ablation analysis}
\begin{tabular}{|c|c|c|c|c|c|c|c|c|}
\hline
Production  & Anomaly   detection & Log template   extraction & Is clustering & Precision & Cover rate & Time & Num of   nodes & F1 \\ \hline
\multirow{7}{*}{MaxCompute}  & \textbf{ROBUST STL} & \textbf{AFT-TREE} & \textbf{Hierarchical   Clustering} & \textbf{79.8}\% & \textbf{77.8\%} & 3.2min & 224 & \textbf{0.78} \\ \cline{2-9} 
  & \textbf{NONE} & AFT-TREE & Hierarchical   Clustering & 24.0\% & 33.3\% & \textbf{2.7min} & 224 & 0.27 \\ \cline{2-9} 
  & ROBUST STL & \textbf{NONE} & Hierarchical   Clustering & 38.1\% & 33.30\% & 17.3min & 224 & 0.35 \\ \cline{2-9} 
  & ROBUST STL & AFT-TREE & \textbf{NONE} & 19.8\% & 38.9\% & 57min & 724 & 0.26 \\ \hline
\multirow{7}{*}{RealtimeCompute}  & \textbf{ROBUST STL} & \textbf{AFT-TREE} & \textbf{Hierarchical Clustering} & \textbf{76.3\%} & \textbf{72.2\%} & 2.2min & 173 & \textbf{0.74} \\ \cline{2-9} 
  & \textbf{NONE} & AFT-TREE & Hierarchical   Clustering & 30.2\% & 33.3\% & \textbf{1.8min} & 173 & 0.31 \\ \cline{2-9} 
  & ROBUST STL & \textbf{NONE} & Hierarchical   Clustering & 36.5\% & 33.30\% & 18.1min & 173 & 0.34 \\ \cline{2-9} 
  & ROBUST STL & AFT-TREE & \textbf{NONE} & 23.9\% & 33.3\% & 46min & 651 & 0.27 \\ \hline
\multirow{7}{*}{Hologres}   & \textbf{ROBUST STL} & \textbf{AFT-TREE} & \textbf{Hierarchical   Clustering} & \textbf{60\%} & \textbf{72.2\%} & 1.7min & 154 & \textbf{0.65} \\ \cline{2-9} 
  & \textbf{NONE} & AFT-TREE & Hierarchical   Clustering & 30.2\% & 50\% & 2.1min & 154 & 0.37 \\ \cline{2-9} 
  & ROBUST STL & \textbf{NONE} & Hierarchical   Clustering & 36.5\% & 33.30\% & 14.7min & 154 & 0.34 \\ \cline{2-9} 
  & ROBUST STL & AFT-TREE & \textbf{NONE} & 16.7\% & 50\% & 37min & 584 & 0.25 \\ \hline
\end{tabular}
\label{Ablation analysis}
\end{table*}

The results in Table \ref{Ablation analysis} show that skipping any of the steps in the feature engineering workflow compromises the performance on all the platforms. The performance gain from anomaly detection, log template extraction and log clustering can be explained by fact that our feature engineering flow extracts the most relevant information from complex raw data and thus boosts the efficiency of KHBN in both graph construction and root cause inference.

\subsubsection{Comparison of Different Techniques in Feature Engineering}

In this part of our experiments, a benchmark method is used in place of the proposed at each stage of feature engineering. Specifically, for time series decomposition, we compare Robust STL with traditional STL~\cite{robert1990stl}, and for metric anomaly detection we compare the technique we proposed with Isolation Forest (iForest), an unsupervised detection algorithm proposed by \cite{liu2008isolation}. We compare AFT-tree with FT-tree~\cite{zhang2018prefix} in extracting log templates. As for log clustering, we use DBSCAN~\cite{ester1996density} as a surrogate for hierarchical clustering. 

\begin{table*}[h]
\footnotesize
\caption{Performance of different methods in feature engineering}
\begin{tabular}{|c|c|c|c|c|c|c|c|c|}
\hline
Production  & Anomaly   detection & Log template   extraction & Is clustering & Precision & Cover rate & Time & Num of   nodes & F1 \\ \hline
\multirow{5}{*}{MaxCompute}  & \textbf{ROBUST STL} & \textbf{AFT-TREE} & \textbf{Hierarchical   Clustering} & \textbf{79.8}\% & \textbf{77.8\%} & 3.2min & 224 & \textbf{0.78} \\ \cline{2-9} 
  & \textbf{STL} & AFT-TREE & Hierarchical   Clustering & 58.4\% & 61.1\% & 5.3min & 224 & 0.60 \\ \cline{2-9} 
  & \textbf{iForest} & AFT-TREE & Hierarchical   Clustering & 48.9\% & 50\% & 4.7min & 224 & 0.49 \\ \cline{2-9} 
  & ROBUST STL & \textbf{FT-TREE} & Hierarchical   Clustering & 55.2\% & 61.1\% & 83.2min & 224 & 0.58 \\ \cline{2-9} 
  & ROBUST STL & AFT-TREE & \textbf{DBSCAN} & 72.4\% & 77.80\% & 3.3min & 224 & 0.75 \\ \hline
\multirow{5}{*}{RealtimeCompute}  & \textbf{ROBUST STL} & \textbf{AFT-TREE} & \textbf{Hierarchical Clustering} & \textbf{76.3\%} & \textbf{72.2\%} & 2.2min & 173 & \textbf{0.74} \\ \cline{2-9} 
& \textbf{STL} & AFT-TREE & Hierarchical   Clustering & 48.9\% & 61.1\% & 4.2min & 173 & 0.54 \\ \cline{2-9} 
  & \textbf{iForest} & AFT-TREE & Hierarchical   Clustering & 37.5\% & 61.1\% & 3.5min & 173 & 0.46 \\ \cline{2-9} 
  & ROBUST STL & \textbf{FT-TREE} & Hierarchical   Clustering & 63.5\% & \textbf{72.2}\% & 61.6min & 173 & 0.67 \\ \cline{2-9} 
  & ROBUST STL & AFT-TREE & \textbf{DBSCAN} & 73.2\% & 72.2\% & 2.4min & 173 & 0.72 \\ \hline
\multirow{5}{*}{Hologres}   & \textbf{ROBUST STL} & \textbf{AFT-TREE} & \textbf{Hierarchical   Clustering} & \textbf{60\%} & \textbf{72.2\%} & 1.7min & 154 & \textbf{0.65} \\ \cline{2-9} 
  & \textbf{STL} & AFT-TREE & Hierarchical   Clustering & 55.2\% & 61.1\% & 3.4min & 154 & 0.58 \\ \cline{2-9} 
  & \textbf{iForest} & AFT-TREE & Hierarchical   Clustering & 42.7\% & 61.1\% & 2.9min & 154 & 0.50 \\ \cline{2-9} 
  & ROBUST STL & \textbf{FT-TREE} & Hierarchical   Clustering & 43.8\% & \textbf{72.2}\% & 42.5min & 154 & 0.54 \\ \cline{2-9} 
  & ROBUST STL & AFT-TREE & \textbf{DBSCAN} & 58.4\% & 72.2\% & \textbf{1.5min} & 154 & 0.64 \\ \hline
\end{tabular}
\label{Performance of different methods in feature engineering}
\end{table*}

Table \ref{Performance of different methods in feature engineering} shows that (1) Robust STL leads to significant improvement over traditional STL in both performance and efficiency; (2) The anomaly detection technique we proposed outperforms iForest in all the metrics on all the platforms; (3) AFT-tree exhibits higher computational efficiency than FT-tree, thanks to the improved storage structure and pruning strategy; (4) As for log clustering, DBSCAN and hierarchical clustering perform similarly, indicating that other decent clustering methods may be employed as a substitute of hierarchical clustering.

\subsubsection{Evaluation of KHBN}\label{sec:evavluation of khbn}

In this part of our experiments, we aim to compare the proposed root cause analysis algorithm KHBN with other well studied methods, including CloudRanger, OM knowledge graph based method (OM Graph for short) as well as LogCluster. CloudRanger takes advantage of a dynamic causal relationship analysis approach to construct impact graphs amongst applications without topology knowledge, and then uses a heuristic algorithm based on second-order random walk to identify the root cause metrics on the graph. OM Graph is an approach proposed by Juan Qiu etc. for mining causality and diagnosing root causes that uses knowledge graph technology and a causal search algorithm. LogCluster is an approach that clusters the logs and extracts representative log sequences from the clusters to identify a problem. A comparison of the four methods is summarized in Table \ref{Comparison among different root cause analysis methods}. Note that KHBN, OM Graph and CloudRanger are all graph-based models with different construction principles and inference methodologies. For OM Graph and CloudRanger, the inferred root causes are the services on the graph which in our case correspond to the engineered features, and the precision are labeled by experienced SREs of the cloud platforms. We conduct comparative tests using the four methods on all the three cloud computing platforms and the results are listed in Table \ref{Performance of the four root cause analysis methods}.

\begin{table*}[h]
\footnotesize
\caption{Comparison among different root cause analysis methods}
\begin{tabular}{|c|c|c|c|c|}
\hline
Method & KHBN & LogCluster & CloudRanger & OM Graph \\ \hline
knowledge-based & Yes & No & No & Yes \\ \hline
Graph-based & Yes & No & Yes & Yes \\ \hline
Nodes in the Graph & \begin{tabular}[c]{@{}c@{}}KPIs, Log patterns, \\ root cause module and types\end{tabular}  & - & KPIs & KPIs \\ \hline
graph construction method & PC algorithm & - & causal  analysis & \begin{tabular}[c]{@{}c@{}}causal analysis  and \\ optimized PC algorithm\end{tabular} \\ \hline
inference method & exact inference & Clustering & Random walk & BFS algorithm \\ \hline
inference result & \begin{tabular}[c]{@{}c@{}}the most likely root cause \\ module and types\end{tabular} & representative log sequences & top k candidates of root cause & candidate root cause paths \\ \hline
\end{tabular}
\label{Comparison among different root cause analysis methods}
\end{table*}

\begin{table}[h]
\footnotesize
\caption{Performance of the four root cause analysis methods}
\begin{tabular}{|c|c|c|c|c|c|}
\hline
Production &  \begin{tabular}[c]{@{}c@{}} Framework of \\ RCA\end{tabular} & Precision & \begin{tabular}[c]{@{}c@{}} Cover \\ rate\end{tabular} & F1 & Time \\ \hline
\multirow{4}{*}{\begin{tabular}[c]{@{}c@{}} Max- \\ Compute\end{tabular}}  & KHBN & \textbf{79.8\%} & \textbf{77.8\%} & \textbf{0.78} & 13.5s \\ \cline{2-6} 
  & LogCluster & 44.7\% & 38.9\% & 0.41 & \textbf{7.4s} \\ \cline{2-6} 
  & CLOUDRANGER & 63.5\% & 61.1\% & 0.62 & 17.3s \\ \cline{2-6} 
  & OM GRAPH & 61.3\% & 33.3\% & 0.43 & 15.4s \\ \hline
\multirow{4}{*}{\begin{tabular}[c]{@{}c@{}} Realtime-\\Compute\end{tabular}}  & KHBN & \textbf{76.3\%} & \textbf{72.2\%} & \textbf{0.74} & 11.2s \\ \cline{2-6} 
  & LogCluster & 43.8\% & 33.3\% & 0.37 & \textbf{6.3s} \\ \cline{2-6} 
  & CLOUDRANGER & 70.8\% & 61.1\% & 0.65 & 12.8s \\ \cline{2-6} 
  & OM GRAPH & 66.7\% & 38.9\% & 0.49 & 13.6s \\ \hline
\multirow{4}{*}{Hologres}  & KHBN & \textbf{60\%} & \textbf{72.2\%} & \textbf{0.65} & 9.7s \\ \cline{2-6} 
  & LogCluster & 50\% & 50\% & 0.5 & \textbf{5.7s} \\ \cline{2-6} 
 & CLOUDRANGER & 44.5\% & 33.3\% & 0.38 & 11.9s \\ \cline{2-6} 
  & OM GRAPH & 58.7\% & 38.9\% & 0.46 & 13.7s \\ \hline
\end{tabular}
\label{Performance of the four root cause analysis methods}
\end{table}

The experimental results show that KHBN significantly outperforms LogCluster on all the three platforms. The reason lies in that LogCluster only considers logs and neglects information from KPIs as well as topology. KHBN outperforms CloudRanger and OM Graph in all the three platforms for at least 0.09 increase in f1 score. Among the three platforms, MaxCompute has the highest f1 score which reaches 0.78 because it has the richest historical data for training. We also find that KHBN is the fastest one in building the network model and inferring the root causes, since it's computation is designed to be parallel.

\subsubsection{The Impact of Predefined Knowledge}

As we know from Table \ref{Performance of the models having pre-defined knowledge}, both KHBN and OM Graph are built based on predefined knowledge which is the topology information from CMDB. In real-world cloud systems, however, it may not be easy to acquire complete topology knowledge to build the graph. Therefore we design experiments to study the impact of predefined knowledge on the performance of the four root cause analysis methods mentioned in Section \ref{sec:evavluation of khbn} by eliminating some pre-known relationships between metrics and modules. The results are displayed in Table \ref{Performance of the models having pre-defined knowledge}, which indicates that the f1 score of KHBN are lowered slightly when deployed in Realtime Compute and Hologres, while the performance of OM Graph deteriorates significantly. However, when it comes to MaxCompute, which has adequate training data, both of these two methods have robust performance in identifying the root cause. Therefore, eliminating predefined knowledge in the stage of constructing the model may adversely affect the performance, but the performance drop can be partially offset by the availability of sufficient training data.

\begin{table}[h]
\footnotesize
\caption{Performance of the models having predefined knowledge}
\begin{tabular}{|c|c|c|c|c|c|c|}
\hline
Production  & \begin{tabular}[c]{@{}c@{}}Predefined \\ knowledge\end{tabular} & \begin{tabular}[c]{@{}c@{}}Framework \\ RCA\end{tabular} & Precision & \begin{tabular}[c]{@{}c@{}}Cover \\ rate\end{tabular} & F1 & Time \\ \hline
\multirow{4}{*}{\begin{tabular}[c]{@{}c@{}} Max- \\ Compute\end{tabular}} &  CMDB & KHBN & \textbf{79.8\%} & \textbf{77.8\%} & \textbf{0.78} & \textbf{13.5s} \\ \cline{2-7} 
 & NONE & KHBN & 63.3\% & 68.2\% & 0.66 & 29.3s \\ \cline{2-7} 
 &  NONE & \begin{tabular}[c]{@{}c@{}} OM \\ GRAPH\end{tabular} & 32.5\% & 45.7\% & 0.37 & 32.4s \\ \hline
\multirow{4}{*}{\begin{tabular}[c]{@{}c@{}} Realtime-\\Compute\end{tabular}}  & CMDB & KHBN & \textbf{76.3\%} & \textbf{72.2\%} & \textbf{0.74} & \textbf{11.2s} \\ \cline{2-7} 
 &  NONE & KHBN & 63.3\% & 38.1\% & 0.47 & 25.1s \\ \cline{2-7} 
 &  NONE & \begin{tabular}[c]{@{}c@{}} OM \\ GRAPH\end{tabular} & 34.3\% & 42.3\% & 0.37 & 27.8s \\ \hline
\multirow{4}{*}{Hologres} &  CMDB & KHBN & \textbf{60\%} & \textbf{72.2\%} & \textbf{0.65} & \textbf{9.7s} \\ \cline{2-7} 
 &  NONE & KHBN & 49.3\% & 68.3\% & 0.57 & 22.5s \\ \cline{2-7} 
 &  NONE & \begin{tabular}[c]{@{}c@{}} OM \\ GRAPH\end{tabular} & 29.7\% & 16.7\% & 0.21 & 28.8s \\ \hline
\end{tabular}
\label{Performance of the models having pre-defined knowledge}
\end{table}


\subsection{Localization of Novel Root Causes}

With the fast growing scale and rapid development of system architectures, cloud computing platforms often have to cope with new emerging types of root causes. To evaluate the robustness of our models in dealing with new types of root causes, we select several types of root causes from different modules, and exclude those samples related to these selected types from the training set. We use these samples as the test set in our experiments and see whether the model can correctly identify the root cause on the metric or module level. Results are shown in Table \ref{Performance of the models dealing with new types of root causes}. We find that KHBN has the highest f1 score (all above 0.5) in the three products compared with other methods, which is owing to the design of KHBN's hierarchical root cause layer. When new types occur, even though the new type does not exist on the graph, KHBN is able to locate the node in the second layer which indicates the module that the root cause is associated with. In contrast, CloudRanger and OM Graph suffer significant accuracy loss in identifying the specific metric node in the graph, since there may exist strong correlation among those metrics. LogCluster seems to perform similarly in identifying novel and existing root causes.

\begin{table}[h]
\footnotesize
\caption{Performance of the models dealing with new types of root causes}
\begin{tabular}{|c|c|c|c|c|}
\hline
Production  & Framework of   RCA & Precision & Cover rate & F1 \\ \hline
\multirow{4}{*}{\begin{tabular}[c]{@{}c@{}} Max-\\Compute\end{tabular}
}  & KHBN & \textbf{71.2\%} & \textbf{83.3\%} & \textbf{0.76} \\ \cline{2-5} 
  & LogCluster & 22.5\% & 33.3\% & 0.26 \\ \cline{2-5} 
  & CLOUDRANGER & 58.8\% & 16.7\% & 0.26 \\ \cline{2-5} 
  & OM GRAPH & 41.7\% & 50\% & 0.45 \\ \hline
\multirow{4}{*}{\begin{tabular}[c]{@{}c@{}} Realtime-\\Compute\end{tabular}
}  & KHBN & \textbf{56.3\%} & \textbf{66.7\%} & \textbf{0.61} \\ \cline{2-5} 
  & LogCluster & 18.8\% & 16.7\% & 0.17 \\ \cline{2-5} 
  & CLOUDRANGER & 43.8\% & 33.3\% & 0.37 \\ \cline{2-5} 
  & OM GRAPH & 47.3\% & 33.3\% & 0.39 \\ \hline
\multirow{4}{*}{Hologres}  & KHBN & \textbf{43.3\%} & \textbf{66.7\%} & \textbf{0.52} \\ \cline{2-5} 
  & LogCluster & 16.7\% & 33.3\% & 0.22 \\ \cline{2-5} 
  & CLOUDRANGER & 30\% & 33.3\% & 0.31 \\ \cline{2-5} 
  & OM GRAPH & 38.1\% & 66.7\% & 0.48 \\ \hline
\end{tabular}
\label{Performance of the models dealing with new types of root causes}
\end{table}

\subsection{Sensitivity Analysis}

\begin{figure*}[h]
\centerline{\includegraphics[scale=0.4]{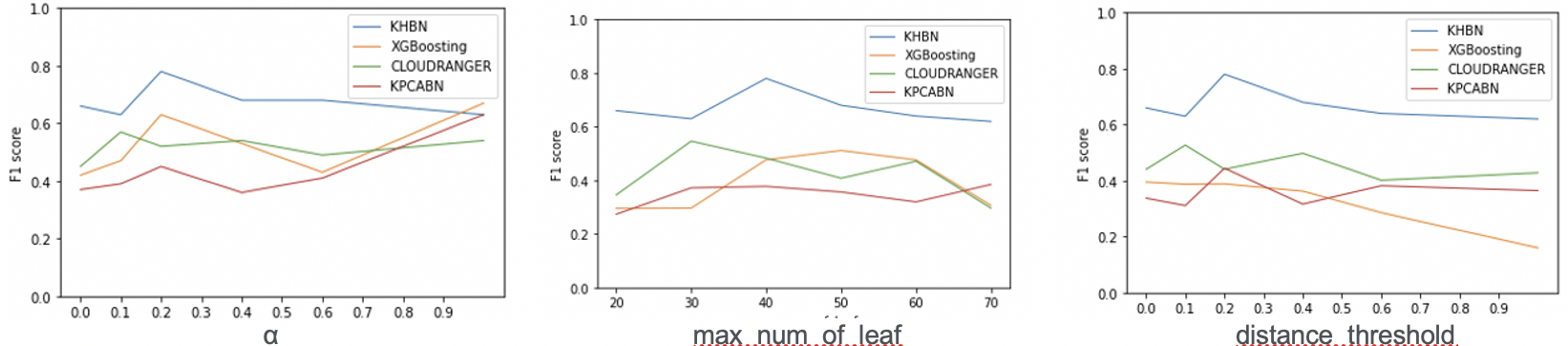}}
\caption{Parameter analysis}
\label{parameters comparing}
\end{figure*}

There are several parameters in our CloudRCA workflow, such as the significance level $\alpha$ which controls the sensitivity of anomaly detection, max\_num\_of\_leaf which affects the log template extraction, and distance\_threshold $\theta$ which controls the number of clusters in log clustering. To study the performance sensitivity in various algorithmic parameters, we conduct experiments by changing one parameter while keeping the others constant. Experiments are performed on the data from MaxCompute. Results of the sensitivity experiments are presented in Figure \ref{parameters comparing}. We see that KHBN consistently outperforms (all above $0.6$) other root cause analysis methods in f1-score across almost all parameter configurations, further demonstrating its superior performance against other methods. KHBN also seems to perform less sensitively with respect to the parameters, and thus can deliver a more stable performance in practice.

\subsection{Computational Scalability}

Based on the data size in Table \ref{The Datasets of the three cloud computing platforms} and number of nodes in Table \ref{Performance of different methods in feature engineering}, we have found that CloudRCA works efficiently in all the three platforms regardless of their different scale. To further demonstrate the scalability of CloudRCA, we evaluate the overheads of our method with synthetic data. We measured the run times of our method when trained with data of different size ranging from 50 to 1000, and different numbers of nodes for  50, 100, 300, 500 and 1000. The results in Table \ref{Overheads of the experiment} indicate that the run time of KHBN linearly increases with the volume of training data and feature size thanks to the parallel computation and that KHBN consistently runs faster ($>50\%$) than the two graph-based competitors. In particular, with as many as 2000 samples and 1000 feature nodes, our workflow is able to finish model construction in 3.5 minutes, a much shorter time period than an SREs typically spends in manual troubleshooting.

\begin{table*}[h]
\footnotesize
\caption{Overheads of the experiment}
\begin{tabular}{|c|c|c|c|c|c|c|}
\hline
Production & Framework of   RCA & 50 nodes & 100 nodes & 300 nodes & 500 nodes & 1000 nodes \\ \hline
\multirow{3}{*}{MaxCompute} & KHBN & 7.8s & 17.4s & 41.5s & 102.3s & 212.7s \\ \cline{2-7} 
 & CLOUDRANGER & 22.4s & 51.2s & 104.3s & 251.8s & 515.4s \\ \cline{2-7} 
 & OM GRAPH & 19.8s & 52.3s & 101.8s & 247.5s & 503.8s \\ \hline
\multirow{3}{*}{RealtimeCompute} & KHBN & 6.9s & 14.6s & 37.7s & 94.6s & 195.8s \\ \cline{2-7} 
 & CLOUDRANGER & 19.6s & 44.7s & 95.3s & 237.3s & 503.7s \\ \cline{2-7} 
 & OM GRAPH & 20.5s & 42.3s & 98.7s & 229.6s & 495.6s \\ \hline
\multirow{3}{*}{Hologres} & KHBN & 5.7s & 11.2s & 29.4s & 91.8s & 196.4s \\ \cline{2-7} 
 & CLOUDRANGER & 18.7s & 37.8s & 94.8s & 228.4s & 487.5s \\ \cline{2-7} 
 & OM GRAPH & 17.3s & 36.1s & 92.9s & 215.7s & 496.4s \\ \hline
\end{tabular}
\label{Overheads of the experiment}
\end{table*}

\subsection{Case Studies}\label{sec:case study}

CloudRCA workflow has been deployed in Alibaba's three big data cloud computing platforms, MaxCompute, Realtime Compute and Hologres. The three platforms support the daily operations of Alibaba as well as the world-wide shopping festival ``Double 11'' with more than 100 thousand servers and complex architectures. The approach we proposed has helped SREs reduce the time of resolving a failure by more than $20\%$. When anomalies are detected in the Key Performance Indicators (KPIs) of a cloud computing platforms, SREs would receive a critical alarm together with the root cause type inferred by our RCA framework through dingtalk, an intelligent working platform created by Alibaba Group. By clicking the url in the alarm message, SREs are directed to an interface where all the important features are displayed. They can confirm those pre-defined automatic operations to remediate the system and provide feedback for the predicted root cause, and all these operations would be recorded and displayed on the interface. In the following, we share some real-world cases where CloudRCA greatly facilitates the identification of root causes when anomalies occur in the platforms.

\textbf{Case 1: New types of root causes in MaxCompute.} Due to the complex architecture and fast iteration of MaxCompute, it can be challenging for SREs to locate root cause types. The case is about a bug in the newly released version of the resource scheduler which prohibited some jobs from acquiring enough resources to finish on time. Since the root cause lied in the new version of the system, SREs had never seen this before and had no clues as to where to start troubleshooting. CloudRCA speeded up the procedure significantly by locating the root cause at the resource management module precisely, which helped SREs shrink the scope of debugging and save the time cost from up to 2 hours to a few minutes.

\textbf{Case 2: Multiple anomalies in Realtime Compute.} There was a fault related to a hotspot issue among the hosts of Realtime Compute. Since the hotspot affected the normal scheduler of jobs as well thus several KPIs in various modules became abnormal simultaneously, SREs got confused in locating the true underlying root cause. CloudRCA successfully identified the right root cause because it combines data from multiple sources, including not only KPIs but also logs of various modules. In this way, CloudRCA helped to reduce the time cost of troubleshooting by nearly 50\%. 

\textbf{Case3: Hidden faults in Hologres.} At the early development stage of Hologres, the KPIs in the monitoring systems were very limited. A batch of jobs running on Hologres encountered an unexpected failure, but none of the KPIs in the monitoring systems alarmed. Fortunately, CloudRCA not only detected the anomaly in log patterns but also provides the right root cause, since certain types of error logs kept increasing unexpectedly.

\subsection{Cross-Platform Transfer Learning}
\begin{table*}[h!]
\footnotesize
\caption{Performance for different modules before and after transfer learning}
\begin{tabular}{|c|c|c|c|c|}
\hline
Production & Module & \begin{tabular}[c]{@{}c@{}}Precision \\ before $\rightarrow$ after\end{tabular} & \begin{tabular}[c]{@{}c@{}}Cover rate \\ before $\rightarrow$ after\end{tabular} & \begin{tabular}[c]{@{}c@{}}F1 \\ before $\rightarrow$ after\end{tabular} \\ \hline
\multirow{5}{*}{MaxCompute} & Resource Scheduler & 86.3\% & 100\% & 0.93 \\ \cline{2-5} 
 & Storage & 79.1\% $\rightarrow$ 90.1\% & 66.7\% & 0.72 $\rightarrow$ 0.76 \\ \cline{2-5} 
 & Host & 83.8\% $\rightarrow$ 88.8\% & 60\% & 0.7 $\rightarrow$ 0.72 \\ \cline{2-5} 
 & Network & 67.6\% $\rightarrow$ 87.3\% & 100\% & 0.81 $\rightarrow$ 0.93 \\ \cline{2-5} 
 & Other & 81.1\% & 75\% & 0.78 \\ \hline
\multirow{5}{*}{RealtimeCompute} & Resource Scheduler & 68.8\% & 100\% & 0.82 \\ \cline{2-5} 
 & Storage & 85.7\% $\rightarrow$ 92.9\% & 60\% $\rightarrow$ 100\% & 0.71 $\rightarrow$ 0.96 \\ \cline{2-5} 
 & Host & 78.9\% $\rightarrow$ 84.2\% & 40\% $\rightarrow$ 80\% & 0.53 $\rightarrow$ 0.82 \\ \cline{2-5} 
 & Network & 66.7\% $\rightarrow$ 75.0\% & 100\% & 0.8 $\rightarrow$ 0.86 \\ \cline{2-5} 
 & Other & 78.9\% & 100\% & 0.88 \\ \hline
\multirow{5}{*}{Hologres} & Resource Scheduler & 54.5\% $\rightarrow$ 68.8\% & 100\% & 0.71 $\rightarrow$ 0.82 \\ \cline{2-5} 
 & Storage & 63.6\% $\rightarrow$ 81.8\% & 60\% $\rightarrow$ 100\% & 0.62 $\rightarrow$ 0.9 \\ \cline{2-5} 
 & Host & 59.2\% $\rightarrow$ 78.5\% & 60\% $\rightarrow$ 80\% & 0.6 $\rightarrow$ 0.79 \\ \cline{2-5} 
 & Network & 63.2\% $\rightarrow$ 78.9\% & 66.7\% $\rightarrow$ 100\% & 0.65 $\rightarrow$ 0.88 \\ \cline{2-5} 
 & Other & 56.7\% & 100\% & 0.72 \\ \hline
\end{tabular}
\label{Performance for different modules before and after transfer learning}
\end{table*}

From the above experiments, we can see that models deployed on MaxCompute outperform Hologres significantly in almost every setup, since Hologres is a fairly new product with limited historical training data while MaxCompute is more mature. We notice that MaxCompute, Realtime Compute and Hologres share some common modules in their architecture as shown in Table \ref{System Modules and Corresponding Root Cause Types of the Three Platforms}, since they belong to Alibaba's big data ecosystem and some modules may have the same KPIs. For example, although the three platforms run on different clusters, they share similar features on the Host and Network modules. As cloud has become an infrastructure with ever increasing popularity, more and more new systems are built on cloud, therefore a general mechanism to improve the root cause analysis performance in a relatively new cloud system can be beneficial. We designed a cross-platform transfer learning mechanism, which combines samples of the same modules from three different big data cloud computing platforms, so as to enrich the training set.

The experiment results are shown in Table \ref{Performance for different modules before and after transfer learning}, where we evaluate the performance of KHBN on each module separately. Numbers after the arrow show performance after applying the transfer learning mechanism. We can see that some common modules such as host, networks and storage benefit from transfer learning remarkably, while there is limited improvement in resource scheduler modules and storage module in Realtime Compute, since the techniques underlying these modules are relatively dependent. Another important trend is that cloud systems are developing its architecture toward could-native, and many cloud-native techniques such as Kubernetes has become a de facto standard, therefore transfer learning can be a promising approach to improve root cause analysis in relatively new cloud products.
\subsection{Key Insights}
We have the following key insights based on our comprehensive experiments: (1) Feature engineering lays the foundation for root cause inference in complex cloud systems by effectively extracting the most relevant information and shrinking the complexity of data, which eventually boosts both accuracy and efficiency in RCA. (2) Combining multiple sources of data provides RCA models with more comprehensive information to precisely locate root causes. (3) Predefined Knowledge and rich training data help improve the accuracy of RCA significantly. (4) Transfer learning across platforms can improve model accuracy when only limited training samples are available from relatively new cloud platforms, and can be a promising direction in cloud-native RCA.

\section{Conclusion and Future Works}
As Alibaba's business keeps developing and flourishing, big data cloud computing platforms are faced with ever increasing challenges in providing stable and reliable service. In our study, we propose CloudRCA, a general framework including anomaly detection, log clustering and Bayesian network inference, to help detect anomalies and identify the root causes when systems fail. Comprehensive experiments have been done in our research to demonstrate the superiority, scalability and robustness of CloudRCA compared with other well-studied RCA frameworks. Moreover, CloudRCA has been deployed in Alibaba's three typical cloud computing platforms including MaxCompute, Realtime Compute and Hologres and helps operators reduce the time of resolving a failure by more than 20\% in the past 12 months. In addition, transfer learning across different platforms can improve the accuracy by more than 10\% , which is promising for Alibaba's big data ecosystem in the upcoming cloud native era. As a supplement and improvement, we'll try GAN together with active learning in the future to reduce the cost of manual fault injection and labeling.

\bibliographystyle{ACM-Reference-Format}
\balance
\bibliography{RCA_reference}


\begin{thebibliography}{25}


\ifx \showCODEN    \undefined \def \showCODEN     #1{\unskip}     \fi
\ifx \showDOI      \undefined \def \showDOI       #1{#1}\fi
\ifx \showISBNx    \undefined \def \showISBNx     #1{\unskip}     \fi
\ifx \showISBNxiii \undefined \def \showISBNxiii  #1{\unskip}     \fi
\ifx \showISSN     \undefined \def \showISSN      #1{\unskip}     \fi
\ifx \showLCCN     \undefined \def \showLCCN      #1{\unskip}     \fi
\ifx \shownote     \undefined \def \shownote      #1{#1}          \fi
\ifx \showarticletitle \undefined \def \showarticletitle #1{#1}   \fi
\ifx \showURL      \undefined \def \showURL       {\relax}        \fi
\providecommand\bibfield[2]{#2}
\providecommand\bibinfo[2]{#2}
\providecommand\natexlab[1]{#1}
\providecommand\showeprint[2][]{arXiv:#2}

\bibitem[\protect\citeauthoryear{Ali, Fakhri, Hayati, Ramli, and Jusoh}{Ali
  et~al\mbox{.}}{2017}]%
        {ali2017n}
\bibfield{author}{\bibinfo{person}{Mohd Rivaie~Mohd Ali},
  \bibinfo{person}{Muhammad~Imza Fakhri}, \bibinfo{person}{Nujma Hayati},
  \bibinfo{person}{Nurul~Atikah Ramli}, {and} \bibinfo{person}{Ibrahim Jusoh}.}
  \bibinfo{year}{2017}\natexlab{}.
\newblock \showarticletitle{The n-th section method: A modification of
  Bisection}.
\newblock \bibinfo{journal}{\emph{Malaysian Journal of Fundamental and Applied
  Sciences}} \bibinfo{volume}{13}, \bibinfo{number}{4} (\bibinfo{year}{2017}),
  \bibinfo{pages}{728--731}.
\newblock


\bibitem[\protect\citeauthoryear{Ester, Kriegel, Sander, Xu,
  et~al\mbox{.}}{Ester et~al\mbox{.}}{1996}]%
        {ester1996density}
\bibfield{author}{\bibinfo{person}{Martin Ester}, \bibinfo{person}{Hans-Peter
  Kriegel}, \bibinfo{person}{J{\"o}rg Sander}, \bibinfo{person}{Xiaowei Xu},
  {et~al\mbox{.}}} \bibinfo{year}{1996}\natexlab{}.
\newblock \showarticletitle{A density-based algorithm for discovering clusters
  in large spatial databases with noise.}. In \bibinfo{booktitle}{\emph{Kdd}},
  Vol.~\bibinfo{volume}{96}. \bibinfo{pages}{226--231}.
\newblock


\bibitem[\protect\citeauthoryear{Gower and Ross}{Gower and Ross}{1969}]%
        {gower1969minimum}
\bibfield{author}{\bibinfo{person}{John~C Gower} {and}
  \bibinfo{person}{Gavin~JS Ross}.} \bibinfo{year}{1969}\natexlab{}.
\newblock \showarticletitle{Minimum spanning trees and single linkage cluster
  analysis}.
\newblock \bibinfo{journal}{\emph{Journal of the Royal Statistical Society:
  Series C (Applied Statistics)}} \bibinfo{volume}{18}, \bibinfo{number}{1}
  (\bibinfo{year}{1969}), \bibinfo{pages}{54--64}.
\newblock


\bibitem[\protect\citeauthoryear{Hochenbaum, Vallis, and Kejariwal}{Hochenbaum
  et~al\mbox{.}}{2017}]%
        {hochenbaum2017automatic}
\bibfield{author}{\bibinfo{person}{Jordan Hochenbaum}, \bibinfo{person}{Owen~S
  Vallis}, {and} \bibinfo{person}{Arun Kejariwal}.}
  \bibinfo{year}{2017}\natexlab{}.
\newblock \showarticletitle{Automatic anomaly detection in the cloud via
  statistical learning}.
\newblock \bibinfo{journal}{\emph{arXiv preprint arXiv:1704.07706}}
  (\bibinfo{year}{2017}).
\newblock


\bibitem[\protect\citeauthoryear{Jia, Yang, Chen, Li, Meng, and Xu}{Jia
  et~al\mbox{.}}{2017}]%
        {jia2017logsed}
\bibfield{author}{\bibinfo{person}{Tong Jia}, \bibinfo{person}{Lin Yang},
  \bibinfo{person}{Pengfei Chen}, \bibinfo{person}{Ying Li},
  \bibinfo{person}{Fanjing Meng}, {and} \bibinfo{person}{Jingmin Xu}.}
  \bibinfo{year}{2017}\natexlab{}.
\newblock \showarticletitle{Logsed: Anomaly diagnosis through mining
  time-weighted control flow graph in logs}. In \bibinfo{booktitle}{\emph{2017
  IEEE 10th International Conference on Cloud Computing (CLOUD)}}. IEEE,
  \bibinfo{pages}{447--455}.
\newblock


\bibitem[\protect\citeauthoryear{Leys, Ley, Klein, Bernard, and Licata}{Leys
  et~al\mbox{.}}{2013}]%
        {leys2013detecting}
\bibfield{author}{\bibinfo{person}{Christophe Leys},
  \bibinfo{person}{Christophe Ley}, \bibinfo{person}{Olivier Klein},
  \bibinfo{person}{Philippe Bernard}, {and} \bibinfo{person}{Laurent Licata}.}
  \bibinfo{year}{2013}\natexlab{}.
\newblock \showarticletitle{Detecting outliers: Do not use standard deviation
  around the mean, use absolute deviation around the median}.
\newblock \bibinfo{journal}{\emph{Journal of experimental social psychology}}
  \bibinfo{volume}{49}, \bibinfo{number}{4} (\bibinfo{year}{2013}),
  \bibinfo{pages}{764--766}.
\newblock


\bibitem[\protect\citeauthoryear{Lin, Zhang, Bannazadeh, and Leon-Garcia}{Lin
  et~al\mbox{.}}{2016a}]%
        {lin2016automated}
\bibfield{author}{\bibinfo{person}{Jieyu Lin}, \bibinfo{person}{Qi Zhang},
  \bibinfo{person}{Hadi Bannazadeh}, {and} \bibinfo{person}{Alberto
  Leon-Garcia}.} \bibinfo{year}{2016}\natexlab{a}.
\newblock \showarticletitle{Automated anomaly detection and root cause analysis
  in virtualized cloud infrastructures}. In \bibinfo{booktitle}{\emph{NOMS
  2016-2016 IEEE/IFIP Network Operations and Management Symposium}}. IEEE,
  \bibinfo{pages}{550--556}.
\newblock


\bibitem[\protect\citeauthoryear{Lin, Zhang, Lou, Zhang, and Chen}{Lin
  et~al\mbox{.}}{2016b}]%
        {lin2016log}
\bibfield{author}{\bibinfo{person}{Qingwei Lin}, \bibinfo{person}{Hongyu
  Zhang}, \bibinfo{person}{Jian-Guang Lou}, \bibinfo{person}{Yu Zhang}, {and}
  \bibinfo{person}{Xuewei Chen}.} \bibinfo{year}{2016}\natexlab{b}.
\newblock \showarticletitle{Log clustering based problem identification for
  online service systems}. In \bibinfo{booktitle}{\emph{2016 IEEE/ACM 38th
  International Conference on Software Engineering Companion (ICSE-C)}}. IEEE,
  \bibinfo{pages}{102--111}.
\newblock


\bibitem[\protect\citeauthoryear{Liu, Ting, and Zhou}{Liu
  et~al\mbox{.}}{2008}]%
        {liu2008isolation}
\bibfield{author}{\bibinfo{person}{Fei~Tony Liu}, \bibinfo{person}{Kai~Ming
  Ting}, {and} \bibinfo{person}{Zhi-Hua Zhou}.}
  \bibinfo{year}{2008}\natexlab{}.
\newblock \showarticletitle{Isolation forest}. In
  \bibinfo{booktitle}{\emph{2008 eighth ieee international conference on data
  mining}}. IEEE, \bibinfo{pages}{413--422}.
\newblock


\bibitem[\protect\citeauthoryear{Marvasti, Poghosyan, Harutyunyan, and
  Grigoryan}{Marvasti et~al\mbox{.}}{2013}]%
        {marvasti2013anomaly}
\bibfield{author}{\bibinfo{person}{Mazda~A Marvasti}, \bibinfo{person}{Arnak~V
  Poghosyan}, \bibinfo{person}{Ashot~N Harutyunyan}, {and}
  \bibinfo{person}{Naira~M Grigoryan}.} \bibinfo{year}{2013}\natexlab{}.
\newblock \showarticletitle{An anomaly event correlation engine: Identifying
  root causes, bottlenecks, and black swans in IT environments}.
\newblock \bibinfo{journal}{\emph{VMware Technical Journal}}
  \bibinfo{volume}{2}, \bibinfo{number}{1} (\bibinfo{year}{2013}),
  \bibinfo{pages}{35--45}.
\newblock


\bibitem[\protect\citeauthoryear{Mikolov, Chen, Corrado, and Dean}{Mikolov
  et~al\mbox{.}}{2013a}]%
        {mikolov2013efficient}
\bibfield{author}{\bibinfo{person}{Tomas Mikolov}, \bibinfo{person}{Kai Chen},
  \bibinfo{person}{Greg Corrado}, {and} \bibinfo{person}{Jeffrey Dean}.}
  \bibinfo{year}{2013}\natexlab{a}.
\newblock \showarticletitle{Efficient estimation of word representations in
  vector space}.
\newblock \bibinfo{journal}{\emph{arXiv preprint arXiv:1301.3781}}
  (\bibinfo{year}{2013}).
\newblock


\bibitem[\protect\citeauthoryear{Mikolov, Sutskever, Chen, Corrado, and
  Dean}{Mikolov et~al\mbox{.}}{2013b}]%
        {mikolov2013distributed}
\bibfield{author}{\bibinfo{person}{Tomas Mikolov}, \bibinfo{person}{Ilya
  Sutskever}, \bibinfo{person}{Kai Chen}, \bibinfo{person}{Greg Corrado}, {and}
  \bibinfo{person}{Jeffrey Dean}.} \bibinfo{year}{2013}\natexlab{b}.
\newblock \showarticletitle{Distributed representations of words and phrases
  and their compositionality}.
\newblock \bibinfo{journal}{\emph{arXiv preprint arXiv:1310.4546}}
  (\bibinfo{year}{2013}).
\newblock


\bibitem[\protect\citeauthoryear{Qiu, Du, Yin, Zhang, and Qian}{Qiu
  et~al\mbox{.}}{2020}]%
        {qiu2020causality}
\bibfield{author}{\bibinfo{person}{Juan Qiu}, \bibinfo{person}{Qingfeng Du},
  \bibinfo{person}{Kanglin Yin}, \bibinfo{person}{Shuang-Li Zhang}, {and}
  \bibinfo{person}{Chongshu Qian}.} \bibinfo{year}{2020}\natexlab{}.
\newblock \showarticletitle{A causality mining and knowledge graph based method
  of root cause diagnosis for performance anomaly in cloud applications}.
\newblock \bibinfo{journal}{\emph{Applied Sciences}} \bibinfo{volume}{10},
  \bibinfo{number}{6} (\bibinfo{year}{2020}), \bibinfo{pages}{2166}.
\newblock


\bibitem[\protect\citeauthoryear{Robert, William, and Irma}{Robert
  et~al\mbox{.}}{1990}]%
        {robert1990stl}
\bibfield{author}{\bibinfo{person}{Cleveland Robert}, \bibinfo{person}{C
  William}, {and} \bibinfo{person}{Terpenning Irma}.}
  \bibinfo{year}{1990}\natexlab{}.
\newblock \showarticletitle{STL: A seasonal-trend decomposition procedure based
  on loess}.
\newblock \bibinfo{journal}{\emph{Journal of official statistics}}
  \bibinfo{volume}{6}, \bibinfo{number}{1} (\bibinfo{year}{1990}),
  \bibinfo{pages}{3--73}.
\newblock


\bibitem[\protect\citeauthoryear{Thalheim, Rodrigues, Akkus, Bhatotia, Chen,
  Viswanath, Jiao, and Fetzer}{Thalheim et~al\mbox{.}}{2017}]%
        {thalheim2017sieve}
\bibfield{author}{\bibinfo{person}{J{\"o}rg Thalheim}, \bibinfo{person}{Antonio
  Rodrigues}, \bibinfo{person}{Istemi~Ekin Akkus}, \bibinfo{person}{Pramod
  Bhatotia}, \bibinfo{person}{Ruichuan Chen}, \bibinfo{person}{Bimal
  Viswanath}, \bibinfo{person}{Lei Jiao}, {and} \bibinfo{person}{Christof
  Fetzer}.} \bibinfo{year}{2017}\natexlab{}.
\newblock \showarticletitle{Sieve: Actionable insights from monitored metrics
  in distributed systems}. In \bibinfo{booktitle}{\emph{Proceedings of the 18th
  ACM/IFIP/USENIX Middleware Conference}}. \bibinfo{pages}{14--27}.
\newblock


\bibitem[\protect\citeauthoryear{Wang, Xu, Ma, Lin, Pan, Wang, and Chen}{Wang
  et~al\mbox{.}}{2018}]%
        {wang2018cloudranger}
\bibfield{author}{\bibinfo{person}{Ping Wang}, \bibinfo{person}{Jingmin Xu},
  \bibinfo{person}{Meng Ma}, \bibinfo{person}{Weilan Lin},
  \bibinfo{person}{Disheng Pan}, \bibinfo{person}{Yuan Wang}, {and}
  \bibinfo{person}{Pengfei Chen}.} \bibinfo{year}{2018}\natexlab{}.
\newblock \showarticletitle{Cloudranger: Root cause identification for cloud
  native systems}. In \bibinfo{booktitle}{\emph{2018 18th IEEE/ACM
  International Symposium on Cluster, Cloud and Grid Computing (CCGRID)}}.
  IEEE, \bibinfo{pages}{492--502}.
\newblock


\bibitem[\protect\citeauthoryear{Wen, Gao, Song, Sun, Xu, and Zhu}{Wen
  et~al\mbox{.}}{2019}]%
        {wen2019robuststl}
\bibfield{author}{\bibinfo{person}{Qingsong Wen}, \bibinfo{person}{Jingkun
  Gao}, \bibinfo{person}{Xiaomin Song}, \bibinfo{person}{Liang Sun},
  \bibinfo{person}{Huan Xu}, {and} \bibinfo{person}{Shenghuo Zhu}.}
  \bibinfo{year}{2019}\natexlab{}.
\newblock \showarticletitle{RobustSTL: A robust seasonal-trend decomposition
  algorithm for long time series}. In \bibinfo{booktitle}{\emph{Proceedings of
  the AAAI Conference on Artificial Intelligence}}, Vol.~\bibinfo{volume}{33}.
  \bibinfo{pages}{5409--5416}.
\newblock


\bibitem[\protect\citeauthoryear{Wen, He, Sun, Zhang, Ke, and Xu}{Wen
  et~al\mbox{.}}{2021}]%
        {WenRobustPeriod20}
\bibfield{author}{\bibinfo{person}{Qingsong Wen}, \bibinfo{person}{Kai He},
  \bibinfo{person}{Liang Sun}, \bibinfo{person}{Yingying Zhang},
  \bibinfo{person}{Min Ke}, {and} \bibinfo{person}{Huan Xu}.}
  \bibinfo{year}{2021}\natexlab{}.
\newblock \showarticletitle{{RobustPeriod}: Robust Time-Frequency Mining for
  Multiple Periodicity Detection}. In \bibinfo{booktitle}{\emph{Proceedings of
  the 2021 International Conference on Management of Data (SIGMOD)}}.
  \bibinfo{pages}{2328--2337}.
\newblock


\bibitem[\protect\citeauthoryear{Wen, Zhang, Li, and Sun}{Wen
  et~al\mbox{.}}{2020}]%
        {FastRobustSTL_wen2020}
\bibfield{author}{\bibinfo{person}{Qingsong Wen}, \bibinfo{person}{Zhe Zhang},
  \bibinfo{person}{Yan Li}, {and} \bibinfo{person}{Liang Sun}.}
  \bibinfo{year}{2020}\natexlab{}.
\newblock \showarticletitle{{Fast RobustSTL}: Efficient and Robust
  Seasonal-Trend Decomposition for Time Series with Complex Patterns}. In
  \bibinfo{booktitle}{\emph{Proceedings of the 26th ACM SIGKDD International
  Conference on Knowledge Discovery \& Data Mining (KDD)}}.
  \bibinfo{pages}{2203--2213}.
\newblock


\bibitem[\protect\citeauthoryear{Weng, Wang, Yang, and Yang}{Weng
  et~al\mbox{.}}{2018}]%
        {weng2018root}
\bibfield{author}{\bibinfo{person}{Jianping Weng}, \bibinfo{person}{Jessie~Hui
  Wang}, \bibinfo{person}{Jiahai Yang}, {and} \bibinfo{person}{Yang Yang}.}
  \bibinfo{year}{2018}\natexlab{}.
\newblock \showarticletitle{Root cause analysis of anomalies of multitier
  services in public clouds}.
\newblock \bibinfo{journal}{\emph{IEEE/ACM Transactions on Networking}}
  \bibinfo{volume}{26}, \bibinfo{number}{4} (\bibinfo{year}{2018}),
  \bibinfo{pages}{1646--1659}.
\newblock


\bibitem[\protect\citeauthoryear{Xu, Chen, Yang, Meng, and Wang}{Xu
  et~al\mbox{.}}{2017}]%
        {xu2017logdc}
\bibfield{author}{\bibinfo{person}{Jingmin Xu}, \bibinfo{person}{Pengfei Chen},
  \bibinfo{person}{Lin Yang}, \bibinfo{person}{Fanjing Meng}, {and}
  \bibinfo{person}{Ping Wang}.} \bibinfo{year}{2017}\natexlab{}.
\newblock \showarticletitle{Logdc: Problem diagnosis for declartively-deployed
  cloud applications with log}. In \bibinfo{booktitle}{\emph{2017 IEEE 14th
  International Conference on e-Business Engineering (ICEBE)}}. IEEE,
  \bibinfo{pages}{282--287}.
\newblock


\bibitem[\protect\citeauthoryear{Yang, Wen, Yang, and Sun}{Yang
  et~al\mbox{.}}{2021}]%
        {yang2021robust}
\bibfield{author}{\bibinfo{person}{Linxiao Yang}, \bibinfo{person}{Qingsong
  Wen}, \bibinfo{person}{Bo Yang}, {and} \bibinfo{person}{Liang Sun}.}
  \bibinfo{year}{2021}\natexlab{}.
\newblock \showarticletitle{A Robust and Efficient Multi-Scale Seasonal-Trend
  Decomposition}. In \bibinfo{booktitle}{\emph{ICASSP 2021-2021 IEEE
  International Conference on Acoustics, Speech and Signal Processing
  (ICASSP)}}. IEEE, \bibinfo{pages}{5085--5089}.
\newblock


\bibitem[\protect\citeauthoryear{Zeng, Tang, Li, Shwartz, and Grabarnik}{Zeng
  et~al\mbox{.}}{2014}]%
        {zeng2014mining}
\bibfield{author}{\bibinfo{person}{Chunqiu Zeng}, \bibinfo{person}{Liang Tang},
  \bibinfo{person}{Tao Li}, \bibinfo{person}{Larisa Shwartz}, {and}
  \bibinfo{person}{Genady~Ya Grabarnik}.} \bibinfo{year}{2014}\natexlab{}.
\newblock \showarticletitle{Mining temporal lag from fluctuating events for
  correlation and root cause analysis}. In \bibinfo{booktitle}{\emph{10th
  International Conference on Network and Service Management (CNSM) and
  Workshop}}. IEEE, \bibinfo{pages}{19--27}.
\newblock


\bibitem[\protect\citeauthoryear{Zhang, Liu, Meng, Luo, Bu, Yang, Liang, Pei,
  Xu, Zhang, et~al\mbox{.}}{Zhang et~al\mbox{.}}{2018}]%
        {zhang2018prefix}
\bibfield{author}{\bibinfo{person}{Shenglin Zhang}, \bibinfo{person}{Ying Liu},
  \bibinfo{person}{Weibin Meng}, \bibinfo{person}{Zhiling Luo},
  \bibinfo{person}{Jiahao Bu}, \bibinfo{person}{Sen Yang},
  \bibinfo{person}{Peixian Liang}, \bibinfo{person}{Dan Pei},
  \bibinfo{person}{Jun Xu}, \bibinfo{person}{Yuzhi Zhang}, {et~al\mbox{.}}}
  \bibinfo{year}{2018}\natexlab{}.
\newblock \showarticletitle{Prefix: Switch failure prediction in datacenter
  networks}.
\newblock \bibinfo{journal}{\emph{Proceedings of the ACM on Measurement and
  Analysis of Computing Systems}} \bibinfo{volume}{2}, \bibinfo{number}{1}
  (\bibinfo{year}{2018}), \bibinfo{pages}{1--29}.
\newblock


\bibitem[\protect\citeauthoryear{Zoubir, Koivunen, Chakhchoukh, and
  Muma}{Zoubir et~al\mbox{.}}{2012}]%
        {zoubir2012robust}
\bibfield{author}{\bibinfo{person}{Abdelhak~M Zoubir}, \bibinfo{person}{Visa
  Koivunen}, \bibinfo{person}{Yacine Chakhchoukh}, {and}
  \bibinfo{person}{Michael Muma}.} \bibinfo{year}{2012}\natexlab{}.
\newblock \showarticletitle{Robust estimation in signal processing: A
  tutorial-style treatment of fundamental concepts}.
\newblock \bibinfo{journal}{\emph{IEEE Signal Processing Magazine}}
  \bibinfo{volume}{29}, \bibinfo{number}{4} (\bibinfo{year}{2012}),
  \bibinfo{pages}{61--80}.
\newblock


\end{thebibliography}







\end{document}